\documentclass[10pt,twocolumn]{IEEEtran}
\usepackage{amsmath,amssymb,mathrsfs}
\usepackage{color,array,times}
\usepackage{epsfig,epstopdf,textcomp}
\usepackage{epsfig, subfigure, color}
\usepackage{float, multirow}
\usepackage{amsmath}
\usepackage{graphicx}
\usepackage{epsfig, subfigure, color}
\usepackage{float, multirow}
\usepackage{epstopdf}
\DeclareGraphicsExtensions{.eps}
\usepackage{amssymb}
\usepackage{algorithm}
\usepackage[noend]{algpseudocode}
\usepackage{tikz}
\usetikzlibrary{matrix}
\usepackage{cancel}
\usepackage{amsmath,stackengine}
\stackMath

\begin{document}

\title{Study of Puncturing Techniques for Polar Codes in 5G Cellular and IoT Networks}

\author{Robert M. Oliveira and Rodrigo C. de Lamare
\thanks{Robert M. Oliveira and Rodrigo C. de Lamare - Centre for Telecommunications Studies (CETUC), Pontifical Catholic University of Rio de Janeiro (PUC-Rio), Rio de Janeiro-RJ, Brasil, E-mails: rbtmota@gmail.com e delamare@cetuc.puc-rio.br} }

\maketitle


\begin{abstract}
This paper presents a puncturing technique based on the channel
polarization index for the design of rate-compatible polar codes in
the fifth generation (5G) of wireless systems. The proposed strategy
consists of two steps: we first generate the codeword message; and
then we reduce the length of the codeword based on the channel
polarization index where channels with the smallest channel
polarization indices are punctured. We consider the proposed
punctured polar codes with the successive cancellation (SC) decoder
and the Cyclic Redundancy Check (CRC) aided SC list (CA-SCL) decoder
and punctured bits known to both the encoder and the decoder. The
Polar Spectra (PS) are then used to performance analysis the
puncturing technique. Simulations for 5G scenarios show that the
proposed polar codes perform comparable to Low-Density Parity-Check
(LDPC) codes.

\end{abstract}


\section{Introdution}

Polar codes was proposed by Arikan [1] in 2009 are based on the
phenomenon called channel polarization, are low-complexity codes and
with the SC decoder can achieve channel capacity when the code
length approaches infinity. For codes of short length, however, SC
decoding falls short in providing a reasonable error-correction
performance. The SCL decoding solves this issue by selecting the
codeword from a list of candidates generated by the decoder. When
SCL is assisted with CRC aided code, polar codes demonstrate
competitive performance as compared to state-of-the-art
error-correcting codes for a wide range of code lengths and have
been adopted in the 5G wireless systems as new radio (NR) [2].

A typical construction of conventional polar codes is based on the
Kronecker product, restricted to the lengths $2^l$ $(l = 1,2,...)$.
Polar codes with arbitrary lengths can be obtained by shortening or
puncturing techniques [3], which will be required for 5G scenarios,
where code lengths ranging from 100 to 1920 bits with various rates
will be adopted [4]. Punctured polar codes can be decoded in a
similar way to SC and SCL decoding for conventional polar codes.

An analysis on the definition of puncturing and shortening
techniques can be found in [3]. In this paper we define the
puncturing technique with elimination the channel synthetic
generated in the polarization channel. The resulting codeword is
smaller than the original one, we call the punctured codeword. Some
puncturing methods of polar codes have been proposed in the
literature and evaluated with the SC decoder, where in the encoder
we freeze a punctured bit channel that receives the fixed zero
value, at the decoder, however, it uses a plus infinite LLR for that
punctured codeword bit as it knows its value with full certainty. In
this paper we use the  technique reported by Hoff [5] where the
punctured bit channels are not processed. The work in [6] introduced
the concept of capacity-one puncturing and devised a simple
puncturing method based on the weight of the columns (CW), whereas
the study in \cite{ref7} proposed a search algorithm to jointly
optimize the shortening patterns and the values of the shortened
bits. The study in \cite{ref8} proposes the reversal quasi-uniform
puncturing scheme (RQUP) based on bit reversed permutation.

In this paper, we propose the study of a puncturing technique based
on the channel polarization index for the design of rate-compatible
polar codes \cite{rc_pd}, where the channel polarization index
determines the choice of the punctured channel. In particular, we
describe the design of rate-compatible  polar codes using the
proposed method and its application to 5G scenarios. A brief
analysis of the proposed puncturing method using the Polar Spectra
[8] is carried out along with the evaluation the performance of
design examples via simulations.

This paper is organized as follows. In Section II we describe the system model and problem statement. In Section III we a brief description of the 5G scenarios. In Section IV we provide a brief description on polar codes. In Section V, we propose puncturing techniques. In Section VI, we have puncturing technical analysis. In Section VII, we evaluate our simulation and, finally, Section VIII concludes this paper.

\section{System Model and Problem Statement}

Fig.1 shows a block diagram of the polar coding system considered
in this paper.

\begin{figure}[htb]
\begin{center}
\includegraphics[scale=0.6]{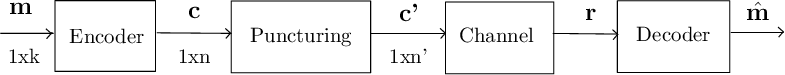} 
\caption{System Model.}
\end{center}
\vspace{-1 em}
\label{figura:fig23b}
\end{figure}

In this system, $\textbf{m}$ is the binary $\textit{message}$
transmitted, with $k$ bits, where $\textbf{m} \in \{0,1\}^{1\rm x
k}$. It is through the generator matrix $\textbf{G}$ that the
message $\textbf{m}$ is encoded, producing the $\textit{codeword}$
with $n$ bits, that is, $\textbf{c} = \textbf{m} \cdot \textbf{G}$
with $\textbf{c} \in \{0,1\}^{1\rm x n}$. With an appropriate
puncturing technique, the codeword $\textbf{c}$ has its length
reduced to $n'$ resulting in the puncturing codeword $\textbf{c'}$,
where $2^{l-1} < n' < 2^l$, where $l$ is an integer that defines the
levels in the polarization tree, $l=\log_2n$. The $\textit{punctured
codeword}$ $\textbf{c'}$ is then transmitted over a channel with
additive white Gaussian noise (AWGN), resulting in the received
vector, $\textbf{r}=\textbf{c'}+\textbf{x}$, where $\textbf{x}$ is
the vector corresponding to the noise. In the decoding step, the
decoding algorithm observes $\textbf{r}$ in order to estimate
$\textbf{m}$. We call it an estimated message $\hat{\textbf{m}}$,
and if $\textbf{m}=\hat{\textbf{m}}$ we say that the message has
been fully recovered. The problem we are interested in solving is
how to design puncturing  with the best code performance.

\section{5G scenarios}

During the 5G standardization process of the 3rd generation
partnership project (3GPP), polar codes have been adopted as channel
coding for uplink and downlink control information for the enhanced
mobile broadband (eMBB) communication service and low-density
parity-check (LDPC) codes have been adopted for the data channels.
5G foresees two other use case, namely ultra-reliable low-latency
communications (URLLC) and massive machine-type communications
(mMTC), for which polar codes have been selected as one of the
possible coding schemes. However, within the 5G framework, various
code lengths, rates and channel conditions are foreseen. Thus,
substantial effort has been put in the design of polar codes that
are easy to implement, having low encoding and decoding complexity,
while maintaining good error-correction performance over multiple
code lengths and channel parameters.

The first scenario called eMBB is dedicated to applications with
high rates of users with different degrees of mobility and rural
environments where transmission rates should be higher than 10Mbps a
10Gbps, latency should be less than 100ms and block size
corresponding to code must be 64800 bits or 1920 bits and the code
rate should be $\rm r = 0.25$ or $\rm r = 0.83$ to ensure high
performance. The second scenario called URLLC is dedicated to
applications IoT and Tactile Internet focuses on transmissions
between machines and Internet of Things devices, where transmission
rates should be in the range of 1Kbps to 10Mbps, the latency should
be between 1 and 100ms, the size of the code word should be 480 bits
and the code rate should be $\rm r = 0.53$. The last scenario,
called mMTC communication, focuses on transmissions between
machines, where transmission rates should be in the range of 1Kbps
to 10Mbps, the latency should be between 1 and 100ms, the size of
the codeword should be 100 bits and the code rate should be $\rm r =
0.64$.

Table I illustrates the requirements mentioned for the different
scenarios of 5G, in which \textit{$R_b$} is the baud rate;
\textit{n} is codeword length; \textit{t} is the latency and $\rm r$
is the coding rate.

\begin{table}[htb]
\caption{\label{tabela}5G requirements}
\begin{center}
\begin{tabular}{|c||c|c|c|}\hline
&eMBB&URLLC&mMTC\\\hline
\textit{$R_b$}&\textit{$R_b$}$>$10Mbps&\textit{$R_b$}$<$10Mbps&\textit{$R_b$}$<$10Mbps\\\hline 
\textit{n}&1920&480&100\\\hline 
\textit{t}&\textit{t}$<$100ms&\textit{t}$<$1ms&\textit{t}$<$100ms\\\hline 
\textit{\rm r}&0.25 or 0.83&0.53&0.64\\ \hline
\end{tabular}
\end{center}
\vspace{-1 em}
\end{table}

\section{Polar Coding System}

Let $W:X \to Y$ denote a binary discrete memoryless channel (B-DMC),
with input alphabet $X=\{0,1\}$, output alphabet $Y$, and the
channel transition probability $W(y|x)$, $x \in X$, $y \in Y$. The
channel mutual information with equiprobable inputs, or symmetric
capacity, is defined by [1]
\begin{equation}
I(W) = \sum\limits_{y \in Y}\sum\limits_{x \in X}\frac{1}{2}W(y|x)log\frac{W(y|x)}{\frac{1}{2}W(y|0)+\frac{1}{2}W(y|1)}
\end{equation}
and the corresponding reliability metric, the Bhattacharyya
parameter is described by [1]
\begin{equation}
Z(W) = Z_0 = \sum\limits_{y \in Y}\sqrt{W(y|0)W(y|1)}
\end{equation}
Applying the channel polarization transform for $n$ independent uses
of $W$, after channel combining and splitting operation we obtain
the group of polarized channels $W_n^{(i)}:X \to Y \times X^{i-1}$,
$i=1,2, \ldots ,n$, defined by the transition probabilities [1]
\begin{equation}
W_n^{(i)}(y_1^n,u_1^{(i-1)}|u_i) = \sum\limits_{u_{i+1}^n \in X^{n-1}}\frac{1}{2^{n-1}}W_n(y_1^n|u_1^n)
\end{equation}
where $(y_1^n,u_1^{i-1})$ denotes the output of $W_n^{(i)}$, $u_i$
its input, where $\textbf{m} = u_1^n$. In general, we use the
notation an $a^n_1$ to designate a vector $(a_1,a_2,...,a_n)$ and
$|a^n_1|$ to refer to its cardinality. The channel polarization
theorem [1] states that $I(W_n^{(i)})$ converges to either 0
(completely noisy channels) or 1 (perfectly noiseless channels) as
$n \to \infty$ and the fraction of noiseless channel tends to
$I(W)$, while polarized channels converge to either $Z(W_n^{(i)})=1$
or $Z(W_n^{(i)})=0$. The vector
$\textbf{m}=(\textbf{u}_A,\textbf{u}_{A^c})$, for some $A \subset
\{1,...,n\}$ denotes the information bit set and $A^c \subset
\{1,...,n\}$ denotes the frozen bit set. We select the
$|\textbf{m}|$ channels to transmit information bits such that
$Z(W_n^{(i)}) \leq Z(W_n^{(j)})$. For encoding, a polar codeword
$\textbf{c}=c^n_1$ is generated by
$\textbf{c}=c^n_1=\textbf{m}\textbf{B}_n\textbf{G}^{\otimes l}_2$,
where $\textbf{m}=u^n_1$ is the information sequence, $\textbf{B}_n$
is the bit-reversal permutation matrix, $\otimes l$ is the $l$-th
Kronecker power and $\textbf{G}_2 =
\footnotesize\left[\begin{array}{cc}
1 & 0 \\
1 & 1 \end{array} \right]$ is the kernel matrix. We adopt the SC
decoder so that the information bits are estimated as [1]
\begin{equation}
\hat{u}_i=\arg\max\limits_{u_i \in \{0,1\}} \ W_n^{(i)}(y_1^n,u_1^{i-1}|u_i),i \in A
\end{equation}

For SCL [2], let $S^{(i)}$ denote the set of candidate sequences in
the $i$th step of the decoding process, and $|S^{(i)}|$ is the size
of $S^{(i)}$. Let $L$ be the maximum allowed size of the list and
$T$ is a threshold for pruning with $T \leq 1$. The SCL algorithm
can be described as follows:
\begin{itemize}
\item bits are estimated successively with index $i = 1, 2, \ldots, n$;
for each candidate in the list,

\item generate two $i$-length
sequences with decoding $\hat{u}_i$ as bit 0 and bit 1 by SC
decoding;

\item if the account of candidates $|S^{(i)}|$ is not larger than $L$,
there is nothing to do;

\item otherwise, reserve $L$ candidates with the largest probabilities
and drop the others from $S^{(i)}$;

\item check each candidate $\hat{u}_i \in S^{(i)}$, if $P(\hat{u}_1^i)
< T \max\limits_{\hat{u}_1^i \in S^{(i)}}P(\hat{u}_1^i)$, eliminate
$\hat{u}_1^i$ from $S^{(i)}$.
\end{itemize}

For sequence determination, after all the bits are examined,
re-encode every candidate in the list and calculate corresponding
likelihood probabilities. Select the one with the maximal
probability as the sequence estimate:
\begin{equation}
\hat{u}_1^n=\arg\max\limits_{\hat{u}_1^n \in S^{(i)}} \prod\limits_{i=1}^{n} W(y_i|u_i=(\hat{u}_1^n)_i).
\end{equation}

The CA-SCL algorithm [2] consists of SCL decoding and CRC decoding
algorithms. The CRC polynomial project [9] depends on the maximum
total length $n$ of the block to be designed considering data and
CRC. According to [10], where $d$ is the degree of the generator
polynomial, the maximum total length of the block is given by
$2^{d-−1}$. The CRC-8 is to be used for $n \leq 128$, whereas the
CRC-16 is suggested for  $n \leq 32768$. Usually, the CRC code is
included in the set of information bits \cite{lbsc}.

\section{Proposed Puncturing Technique Based on Polarization Channel Index}

In this section, we detail the proposed puncturing technique based
on the channel polarization index and show how to calculate the
polarization channels. The purpose of a puncturing technique is to
reduce the codeword length $n$ for $n'$, such that $n' < n$. In
particular, the length reduction is obtained by eliminating bit
channel for codeword with lower polarization indices. Consider the
polarization index vector $\textbf{p}$ which contains the lower
polarization indices, where $|\textbf{p}|=n-n'$ indicates its number
of elements. We firstly expertly generate the codeword by defining
the bits channels in $\textbf{p}$ as a zero value (frozen bits). The
bits of information must be rearranged in the codeword message
according to the order of polarization of the remaining bit
channels. We then reduce the length of the codeword message based on
the channel polarization index, where channels with lower indexes
will be punctured.

The polarization channel $W_n^{(i)}$ is calculated with the
Bhattacharya parameter satisfying the following recursion:
\begin{equation}
\begin{cases}
Z(W^{(2i-1)}_n) = 2Z(W^{(i)}_{n/2})-Z(W^{(i)}_{n/2})^2\\
Z(W^{(2i)}_n) = Z(W^{(i)}_{n/2})^2.\\
\end{cases}
\end{equation}
Using the notation in [1], for $n = 8$ we have $l$ stages of polarization and the $Z_0=0.5$, the polarization stages are
\begin{itemize}
 \item stage 1: \\ $Z(W^+)=2(Z_0)-(Z_0)^2=2(0.5)-(0.5)^2=0.75$ and \\ $Z(W^-)=(Z_0)^2=(0.5)^2=0.25$
 \item stage 2: \\ $Z(W^{++})=2Z(W^+)-Z(W^+)^2=0.937$, \\ $Z(W^{-+})=2Z(W^-)-Z(W^-)^2=0.437$, \\ $Z(W^{+-})=Z(W^+)^2=0.562$ and \\ $Z(W^{--})=Z(W^-)^2=0.062$
 \item stage 3, by induction: \\ $Z(W^{+++})=0.996,\ Z(W^{-++})=0.683,\ Z(W^{+-+})=0.808,\ Z(W^{--+})=0.121,\ Z(W^{++-})=0.878,\ Z(W^{-+-})=0.191,\ Z(W^{+--})=0.316\ \rm and\ Z(W^{---})=0.003$.
\end{itemize}

The channels $(W^{+++},W^{-++},W^{+-+},W^{--+},W^{++-},
\\W^{-+-},W^{+--},W^{---})$ can be written with $(W_0, W_1, W_2,\\
W_3, W_4, W_5, W_6, W_7)$. We define the polarization vector as:
\begin{equation}
\textbf{b} \triangleq \left[\begin{array}{cccc} Z(W_0); Z(W_1); \ldots; Z(W_{n-1})
\end{array}\right]^T
\end{equation}
where the polarization vector $\textbf{b}$ is computed through (6).
As an example for stage 3 we have
\begin{equation}
\textbf{b} = [0.996,0.683,0.808,0.121,0.878,0.191,0.316,0.003]^T.
\end{equation}
The key idea of the proposed technique is to shorten in the codeword $\textbf{m}$ the bits that correspond to the channels with smallest values of polarization index.

These channels can be obtained by sorting the polarization vector
$\textbf{b}$. The goal of sorting is to determine a permutation
$k(1)k(2) \ldots k(n)$ of the indices $\{1,2,...,n\}$ that will
organize the entries of the polarization vector $\textbf{b}$ in
increasing order [11]:
\begin{equation}
Z(W_{k(1)}) \leq Z(W_{k(2)}) \leq \ldots \leq Z(W_{k(n)})
\end{equation}
Consider the sort function $[\textbf{a},\textbf{k}]= \rm
sort(\textbf{b})$ which implements (13), where $\textbf{a}$ lists the
sorted $\textbf{b}$ and $\textbf{k}$ contains the corresponding
indices of $\textbf{a}$. Table II shows an example of the
polarization vector $\textbf{b}$ for $n = 8$, sorting vector
$\textbf{a}$ and the new index $\textbf{k}$.

\begin{table}[htb]
\caption{\label{tabela}Polarization vector $\rm \textbf{b}$ for $n = 8$}
\begin{center}
\vspace{-1 em} \scriptsize
\begin{tabular}{|c||c|c|c|c|c|c|c|c|}\hline
\textit{$\rm
\textbf{b}$}&0.996&0.683&0.808&0.121&0.878&0.191&0.316&0.003\\\hline
\textit{$\rm index$}&1&2&3&4&5&6&7&8\\\hline
\multicolumn{9}{|c|}{After sorting} \\ \hline \textit{$\rm
\textbf{a}$}&0.003&0.121&0.191&0.316&0.683&0.808&0.878&0.996\\\hline
\textit{$\rm \textbf{k}$}&8&4&6&7&2&3&5&1\\\hline
\end{tabular}
\scriptsize \vspace{-0.5 em}
\end{center}
\vspace{-1 em}
\end{table}

The vector $\textbf{k} = [8,4,6,7,2,3,5,1]$ contains the indices of
the polarization values of the channels in increasing order, which
is used to obtain the vector $\textbf{p}$ of the proposed
technique:
\begin{equation}
\textbf{p}=[k(1), \ldots , k(n-n')],
\end{equation}
with $n-n'$ the length of puncturing. In Algorithm 1 we have
included a pseudo-code of the computation of the $\textbf{p}$
vector. Note that $k(\cdot)$ is returned from the sort function.

\begin{algorithm}
\caption{Determine $\textbf{p}$ vector}
\begin{algorithmic}[1]
\State Given a original codeword with length $n$
\State Given a reduced codeword with length $n'$
\State Calculate the polarization channel vector $\textbf{b}$ for $n$
in log domain
\State Calculate $[\textbf{a},\textbf{k}]= \rm sort(\textbf{b})$
\State Calculate the puncturing vector $\textbf{p}=[k(1), \ldots ,
k(n-n')]$
\end{algorithmic}
\end{algorithm}

In Algorithm 2 we have included a pseudo-code of the computation of
the punctured codeword $\textbf{c'}$.

\begin{algorithm}
\caption{Computation of the punctured codeword $\textbf{c'}$}
\begin{algorithmic}[1]
\State Given a reduced codeword with length $\textbf{c}$
\State Given a vector $\textbf{p}$
\State Index each $\textbf{c}$ column by $\{1,2,...,n\}$
\For{y = 1 to $|\textbf{p}|$}
\State $r_{\rm min} \leftarrow \textbf{p}(y)$
\State Delete position from $\textbf{c}$ with index $r_{\rm min}$
\State $\textbf{c'}  \leftarrow \textbf{c}$
\EndFor
\State $\textbf{end for}$
\end{algorithmic}
\end{algorithm}

As the code has been punctured, the reliability of the bit channels
changes and the information set should change accordingly. The
Bhattacharyya parameters of the polarized channels punctured are
smaller than those of the original polarized channels (Lemma 3 in
[8]), we consider in this paper that the order of channel
polarization given by (13) does not change after puncturing. The
punctured codeword $\textbf{c'}$, which contains the bits of the
binary message $\textbf{m}=\textbf{u}_A$ such that $Z(W_{n'}^{(i)})
\leq Z(W_{n'}^{(j)})$ for all $i \in A$, $j \in A^c$ and
$\textbf{u}_{A^c}=(u_i:i \in A^c|u_i=0)$, is then transmitted over a
channel.

We consider now an example with punctured polar codes with length
$n'= 5$. We choose the length of the codeword $\textbf{c}$ as $n=8$.
For the puncturing of $\textbf{c}$ to $\textbf{c'}$, the channels
with the lowest polarization rank values are $W_8$, $W_4$ and $W_6$,
and the puncturing vector is
\begin{equation}
\textbf{p}=(8,4,6)
\end{equation}
and $|\textbf{p}|=3$. For a coding $n=8$ and $k=4$, we have 4 information
bits $(u_1,u_2,u_3,u_4)$. According to the $\textbf{b}$ polarization
vector in (7) we have that the vector $\textbf{m}$
\begin{equation}
\textbf{m}=(0,0,0,u_1,0,u_2,u_3,u_4)
\end{equation}
with the frozen bits set to zero. Given the $\textbf{p}$ puncturing vector in
(11), positions 8, 4 and 6 will be set to zero, and the $\textbf{m}$
bits will be reallocated according to (9). Then the vector
$\textbf{m}$ at the input in the encoder becomes
\begin{equation}
\textbf{m}=(0,u_1,u_2,0,u_3,0,u_4,0).
\end{equation}
According to Algorithm 2, the message encoding $\textbf{c}$ given
by\begin{equation} \textbf{c}=(c_1,c_2,c_3,c_4,c_5,c_6,c_7,c_8),
\end{equation}
must be punctured according to puncturing vector $\textbf{p}$ in
(11), bits 8, 4 and 6 will be eliminated. The $1$st element of $\textbf{p}$ is 8, which
results in the deletion of the $8$th position of $\textbf{c}$. The $2$nd element of $\textbf{p}$ is $4$, which requires the elimination of the $4$th position. At last, the $3$rd element of the $\textbf{p}$ is 6, which requires the deletion of the $6$th position, generating vector\begin{equation}
\textbf{c}'=(c_1,c_2,c_3,c_5,c_7).
\end{equation}

\section{Puncturing Technique Analysis}

An analysis of the performance for polar codes has been presented in
[12] and demonstrates that systematic encoding yields better BER
performance than non-systematic coding with the same FER performance
for both encoding schemes. However, the method used is quite costly
in terms of processing time, since it generates all possible coded
message combinations for a given polar code and calculates the
Hamming distance between them.

An alternative metric called Polar Spectra has been studied in [8],
which has a lower computational cost compared to the method in [12],
and is sufficient to indicate the performance of the adopted
puncturing technique. This metric is based on the channel
polarization tree, the Hamming weight (HW), the complement Hamming
weight (CHW) of each branch and the relation between the polarized
channels $W_i$ and the rows of the generator matrix $\textbf{G}$.

In this paper, the spectrum distance (SD) is considered as the
criterion to compare
puncturing techniques. 
The spectrum distance for path weight (SDP) is given by [8]
\begin{equation}
d = \sum_{(k = 0)}^lP_1(l,k,Q)k = \sum_{(k = 0)}^l\frac{H_n^{(k)}}{n}k,
\end{equation}
where $P_1(l,k,Q)=\frac{H_n^{(k)}}{n}$ is the probability of path
weight $k$ with $|\textbf{p}|$ bits puncturing. The spectrum
distance for the complementary path weight (SDC) is given by [8]
\begin{equation}
\lambda = \sum_{(r = 0)}^lP_0(l,r,Q)r = \sum_{(r = 0)}^l\frac{C_n^{(r)}}{n}r,
\end{equation}
where $P_0(l,k,Q)=\frac{C_n^{(k)}}{n}$. We use SDC as the main
metric to evaluate the performance of the proposed and existing
puncturing techniques.

The SD of the complementary Hamming weight $\lambda$, where
$C_l^{(r)}=\binom{l}{r}$, $l$ is the quantity are the levels in the
polarization tree. The term $C(X)=\sum_{(r = 0)}^lC_l^{(r)}X^r$
describes the total number of branches with a given number of zeros,
or alternatively $C(X)=\sum_{(i=1:n)}X^{\rm Pb_i}$, $\rm Pb$ is the
number of zeros of each branch. As an example, for a
$\textbf{G}_{16}$, we have $C(X)=X^0+4X^1+6X^2+4X^3+X^4$, one branch
with no zero, four branches with 1 zero, six branches with 2 zeros,
4 branches with 3 zeros and one branch with 4 zeros, the $\lambda =
\frac{1 \cdot 0 +4 \cdot 1+6 \cdot 2+4 \cdot 3+1 \cdot 4}{16}=2$.

Given a puncturing, the final metric of the SD is given by $\lambda$
where the $C(X)$ is updated by removing the branches cut by
puncturing, each branch corresponds to a channel, which in turn
corresponds to a row (and column) in the generator matrix
$\textbf{G}$. For $\textbf{G}_{12}$ with $\textbf{p} = (14,15,16)$,
updating $C(X)=2X^1+5X^2+4X^3+1X^4$ and new $\lambda = \frac{2 \cdot
1+5 \cdot 2+4 \cdot 3+1 \cdot 4}{16}=1.75$, always less than the
previous value $\lambda$.

In the Table III we compare the proposed technique with CW[6] and
RQUP[8]. Note that the proposed technique value got has a higher
value than the CW and RQUP techniques.

\begin{table}[htb]
\caption{\label{tabela}Polar Spectra}
\vspace{-0.5 em}
\begin{center}
\begin{tabular}{|c||c|c|c|}\hline
\textit{$\rm SDC$}&Proposed&RQUP&CW\\\hline
\textit{$n'=100,k=64$}&3.73&3.46&3.45\\\hline
\textit{$n'=480,k=256$}&4.62&4.46&4.43\\\hline
\textit{$n'=1920,k=1600$}&5.62&5.44&5.43\\\hline
\end{tabular}
\end{center}
\vspace{-2 em}
\end{table}

\section{Simulation}

In this section, we present simulations of rate-compatible polar
codes with puncturing and a system equipped with the SCL decoder, as
described for the  eMBB 5G scenario, which requires the use of short
to moderate block lengths. We measure the BER against the
signal-to-noise ratio, defined as the ratio of the bit energy,
$E_b$, and the power spectral density, $N_0$, in dB. The algorithm
for polar codes use the Tal-Vardy method [13].

In the first example, we consider the URLLC scenario with $n'=480$
and $k=256$ with the SC decoder. The results  in Fig. 2 show that
the proposed puncturing technique (PD) outperforms the RQUP and the
CW techniques by up to $0.15$dB in $E_b/N_0$ for the same BER
performance, and approaches the performance of the mother code (MC)
with $n=512$ and $k=256$.

\begin{figure}[htb]
\vspace{-1 em}
\begin{center}
\includegraphics[scale=0.54]{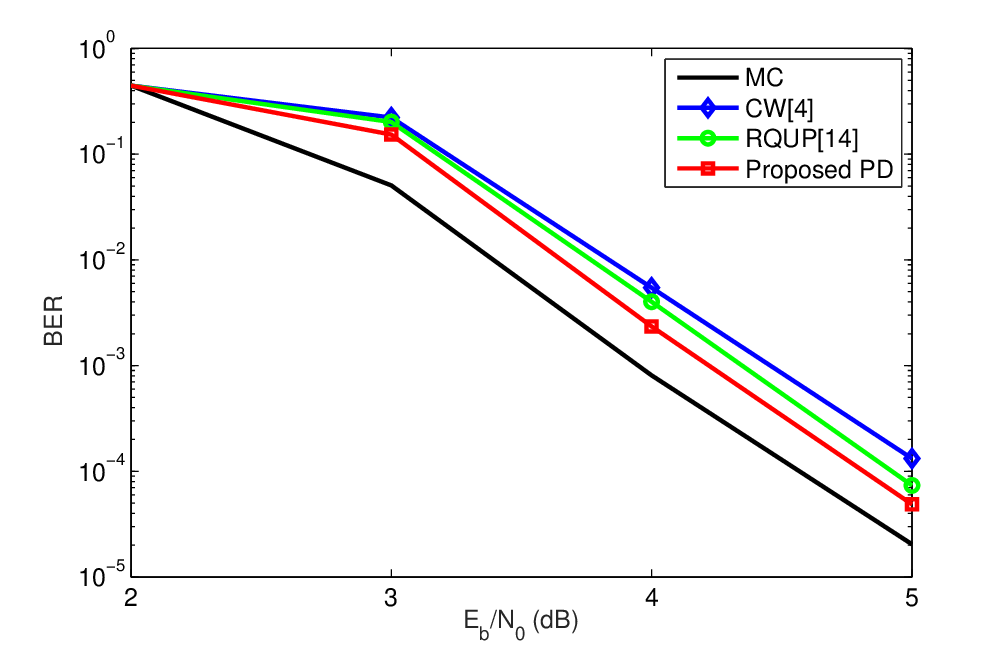}
\caption{BER performace of rate-compatible polar codes $n'$=1920
$k$=1600.}
\end{center}
\label{figura:fig04}
\vspace{-1 em}
\end{figure}

In the second example, we assess the list decoders using different
CRC and with the proposed puncturing technique. Fig. 3 shows the
performance of the list decoders with different CRC for systematic
polar codes in the URLLC scenario with $n'=480$ and $k=256$. The
results show that the performance of the list decoder improves with
the increase of candidates in the lists and the size of the CRC.

\begin{figure}[htb]
\begin{center}
\includegraphics[scale=0.48]{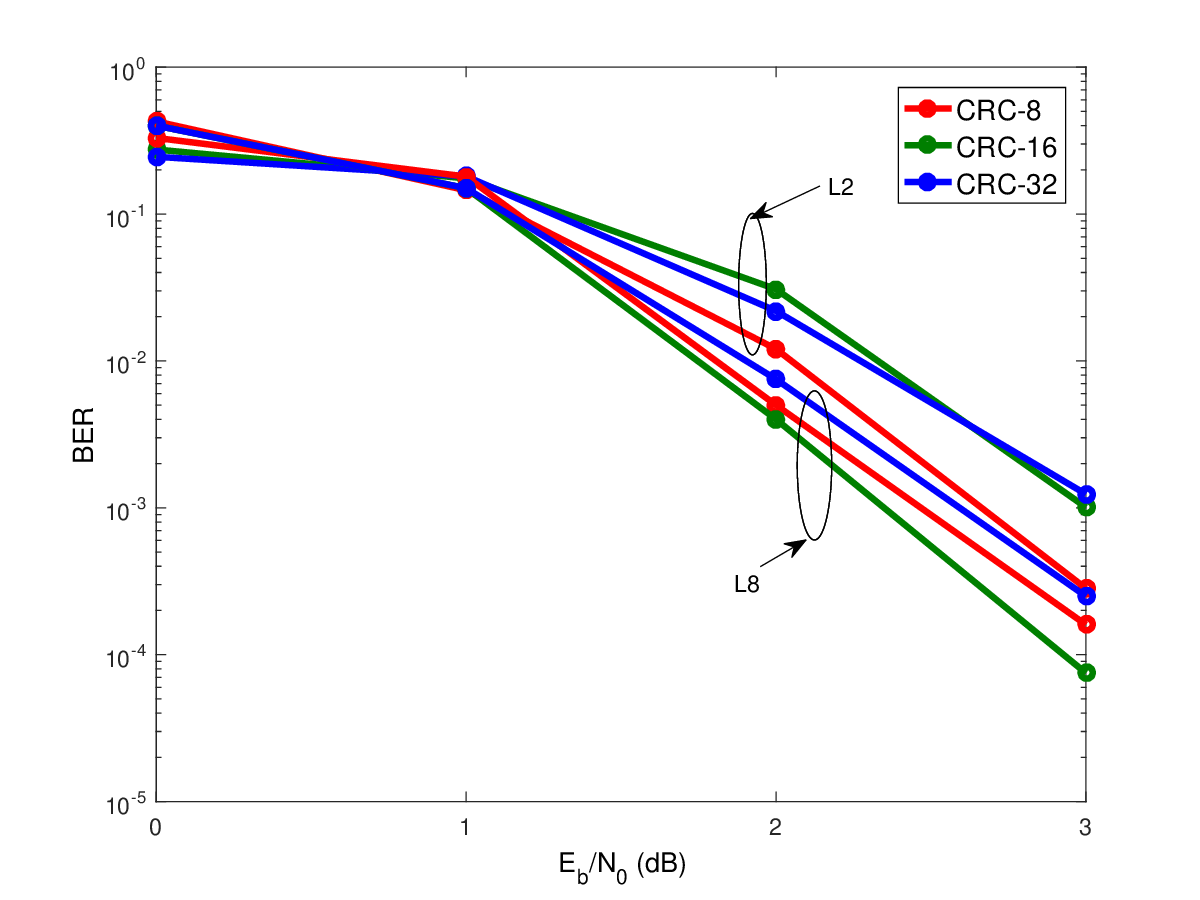}
\caption{Comparison of list decoder with different CRC.}
\end{center}
\label{figura:comparison_CRC_Lista}
\vspace{-2 em}
\end{figure}

In the third example, we consider the eMBB scenario with $n'=1920$
and $k=1600$ and a comparison with LDPC codes designed using a
Progressive Edge Growth (PEG) technique \cite{ref14,dopeg,ref15}.
Even though the LDPC codes are decoded with the standard sum-product
algorithm other decoders such as \cite{vfap} can also be used. The
results in Fig. 4 show that the proposed puncturing technique for
polar codes with SC decoder achieves a similar BER performance to
LDPC codes, which were decoded with the sum-product algorithm using
40 iterations. The results show that LDPC and Polar codes are
comparable in performance.

\begin{figure}[htb]
\vspace{-1 em}
\begin{center}
\includegraphics[scale=0.47]{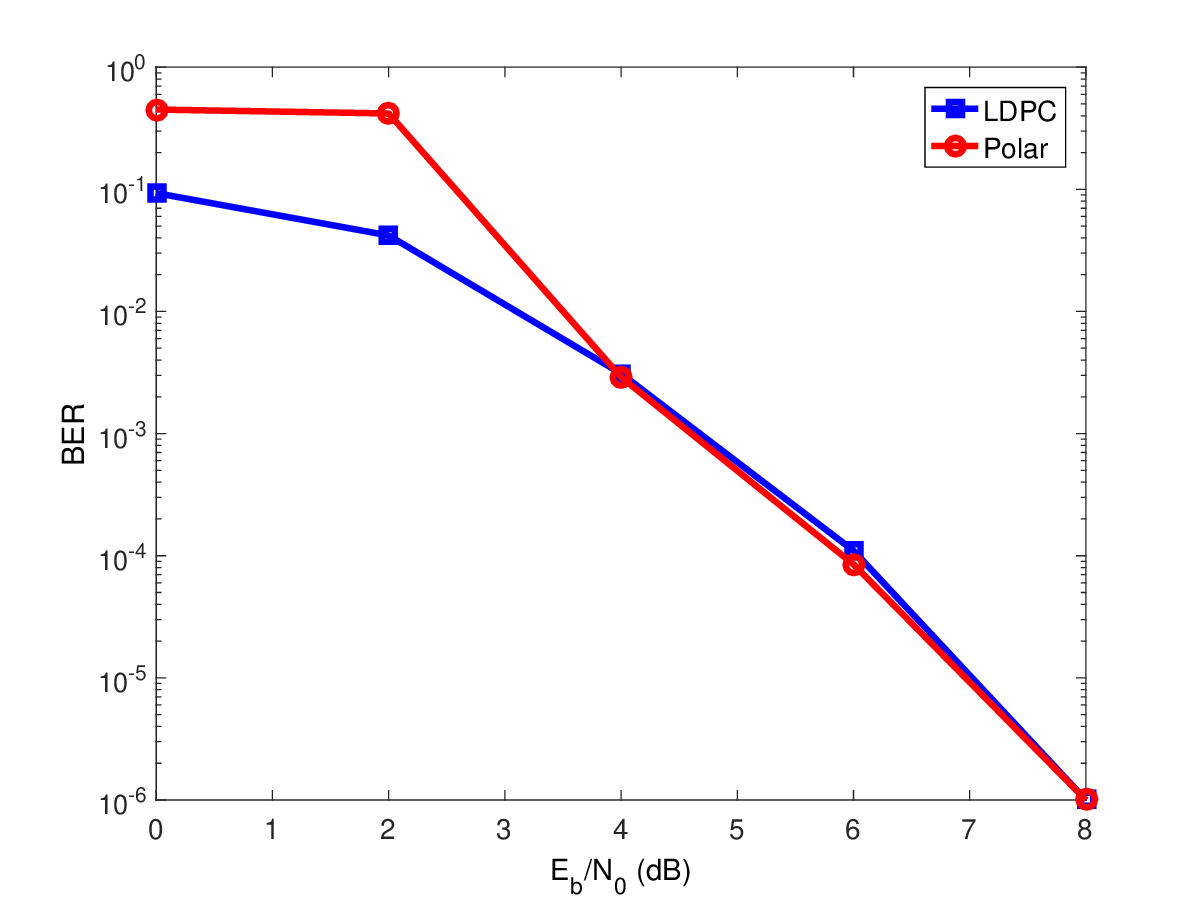}
\caption{BER performace of rate-compatible Polar Code $n'$=1920 $k$=1600.}
\end{center}
\label{figura:fig04}
\vspace{-2 em}
\end{figure}

Future work will consider applications to multiple-antenna systems
\cite{mmimo,wence}, precoding
\cite{sintprec,sintprec2,lcrbd,gbd,wlbd,mbthp,rmbthp,bbprec} and
iterative detection and decoding
\cite{stbcmimo,jio,jidf,smtvb,jiomimo,rrmber,mberdf,spa,mfsic,mfdf,mbdf,did,tds,armo,badstc,1bitidd,baplnc}
techniques.

\section{Conclusion}

We have proposed a puncturing method, which is based on the channel
polarization index, that can bring a performance improvement in
punctured polar codes as compared to existing puncturing methods in
the literature. We have then designed rate-compatible polar codes
for 5G scenarios using the proposed approach and carried out
comparisons with existing puncturing techniques and LDPC codes. The
use of the polar spectra as a benchmark for performance comparison
has also been shown as a valuable tool to indicate the best
puncturing strategy, while requiring a low computational complexity.
Simulations have shown that the performance of the proposed
puncturing technique for the eMBB 5G scenario using the CA-SCL
decoder is quite competitive as compared to existing rate-compatible
polar and LDPC codes.

\end{document}